\documentclass[final,3p,times,twocolumn]{elsarticle}

\usepackage{amssymb}
\usepackage{amsmath}
\usepackage{lineno}
\usepackage{bm}
\usepackage{siunitx}

\journal{Journal of Information Security and Applications}

\begin{document}

\begin{frontmatter}



\title{Keystroke Detection by Exploiting Unintended RF Emission from Repaired USB Keyboards\tnoteref{tlab}}

\tnotetext[tlab]{This research was supported by the Office of the Director of National Intelligence (ODNI), Intelligence Advanced Research Projects Activity (IARPA), via contract: 2021-21062400006.}
\author[Purdue]{Md Faizul Bari\corref{cor1}}
\ead{mbari@purdue.edu}
\author[Purdue]{Yi Xie}
\author[Purdue]{Meghna Roy Chowdhury}
\author[Purdue]{Shreyas Sen}
\cortext[cor1]{Corresponding Author.}
\affiliation[Purdue]{organization={Elmore Family School of Electrical and Computer Engineering, Purdue University},
            city={West Lafayette},
            postcode={47907},
            state={IN},
            country={USA}}



\begin{abstract}
Electronic devices and cables inadvertently emit RF emissions as a byproduct of signal processing and/or transmission. Labeled as electromagnetic emanations, they form an EM side-channel for data leakage. Previously, it was believed that such leakage could be contained within a facility since they are weak signals with a short transmission range. However, in the preliminary version of this work \cite{broken_cable}, we found that the traditional cable repairing process forms a tiny monopole antenna that helps emanations transmit over a long range. Experimentation with three types of cables revealed that emanations from repaired cables remain detectable even at ${>}4$ m and can penetrate a $14$ cm thick concrete wall. In this extended version, we show that such emanation can be exploited at a long distance for information extraction by detecting keystrokes typed on a repaired USB keyboard. By collecting data for 70 different keystrokes at different distances from the target in 3 diverse environments (open space, a corridor outside an office room, and outside a building) and developing an efficient detection algorithm, $\sim$100\% keystroke detection accuracy has been achieved up to 12 m distance, which is the highest reported accuracy at such a long range for USB keyboards in the literature. The effect of two experimental factors, interference and human-body coupling, has been investigated thoroughly. Along with exploring the vulnerability, multi-layer external metal shielding during the repairing process as a possible remedy has been explored. This work exposes a new attack surface caused by hardware modification, its exploitation, and potential countermeasures.
\end{abstract}



\begin{keyword}
TEMPEST \sep Eavesdropping \sep Data Leakage \sep Compromising Emanation \sep Shielding \sep USB \sep Keyboard 

\end{keyword}

\end{frontmatter}



\section{Introduction}
\label{sec:introduction}
\subsection{Background}
\begin{figure}[t]
\centering
\includegraphics[width=0.4\textwidth]{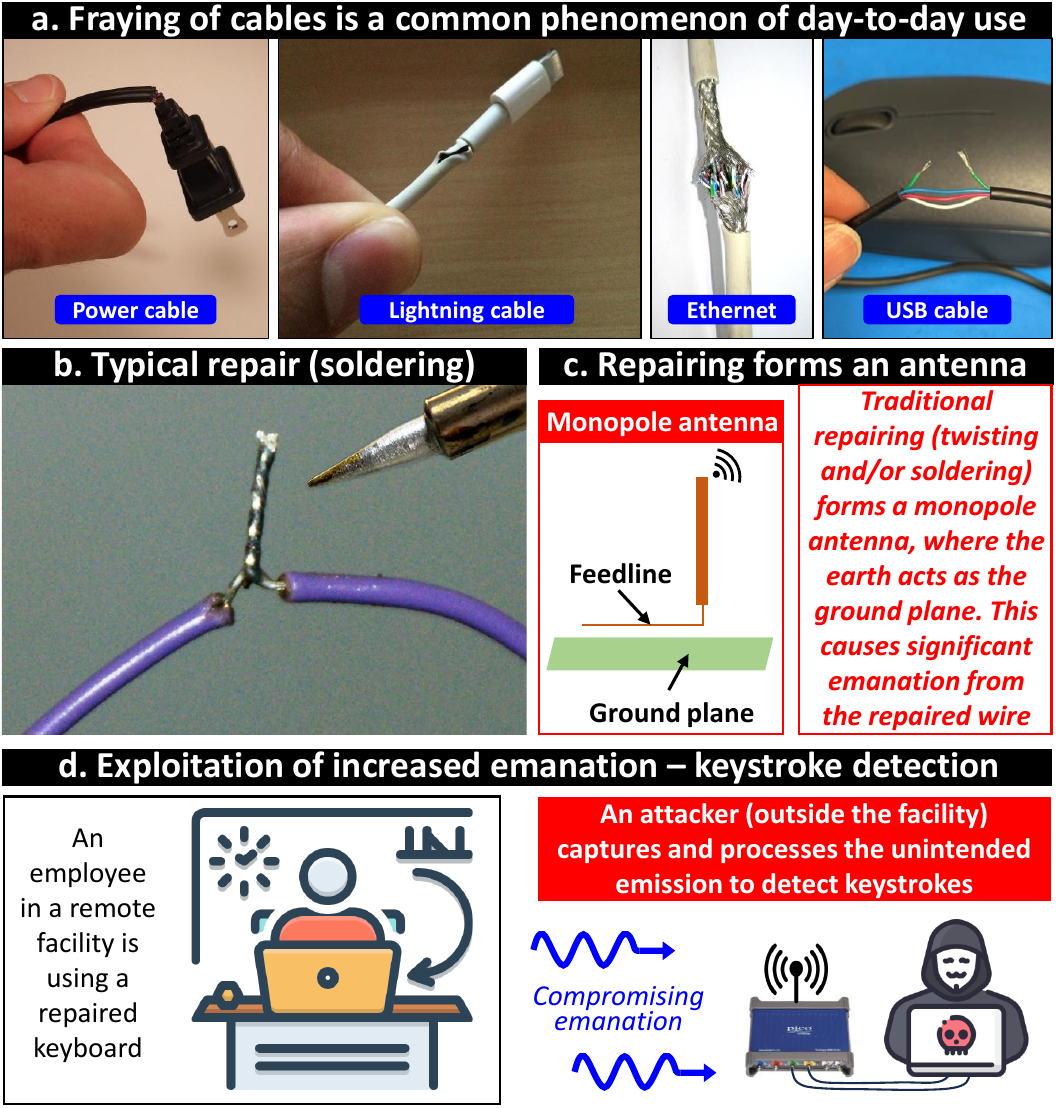}
\caption{(a) The condition of electrical cables deteriorates over day-to-day use and often breaks due to bending, frequent twisting, heat, excessive tensile force, squeezing by heavy objects, etc. It is a common phenomenon, and typically it is much more economical to repair them. (b) Traditional repairing involves twisting to join severed wires or soldering them. (c) These repairing methods form a monopole antenna as a byproduct. As a byproduct of signal carrying, cables already have unintended, weak electromagnetic emissions called EM emanations, which are information leakage sources. Now, the presence of an antenna makes them even stronger. The increased SNR of the emanation renders the whole system vulnerable. (d) The amplified emanation can be exploited for information stealing. One such scenario is an employee within an in-the-wild facility using a repaired USB keyboard while an attacker is capturing the enhanced emanation signal from outside the facility and detecting keystrokes.}
\label{intro}
\end{figure}

Electrical cables are as important as electronic devices to ensure proper power supply and data transfer. However, the quality of insulation and internal conducting material degrades over time, resulting in frayed or broken cables as shown in Fig.~\ref{intro}(a). Other causes behind frayed cables are bending, frequent twisting, heat, excessive tensile force, squeezing by heavy objects (e.g., chair, table, etc.), chewing by pets, etc. In most cases, one or more of the internal wires get broken and can be repaired easily by following three simple steps: (1) cutting off the damaged portion, (2) attaching the conductive part by twisting it together (and soldering, if possible), and (3) adding proper insulation by electrical tapes or heat shrink tubes. While an average consumer has the option to either repair or replace such equipment, in-the-wild facilities are often limited to the repair option. These are government facilities in foreign lands (often in remote regions), and workers in such facilities cannot purchase similar equipment off the local market from an unauthorized vendor due to regulations. Employees in such an in-the-wild facility may have to repair the broken cable to fix the equipment and use it till a new supply arrives. IARPA (Intelligence Advanced Research Projects Activity) has launched the SCISRS program \cite{scisrs} to secure such facilities by detecting anomalous RF signals. This work is a part of that project where we explore the effect of repairing damaged cables in the RF domain.

The traditional repairing methods have two major goals: ensuring electrical connectivity through the wires and insulating the wires properly. However, they ignore the change in the electromagnetic emission from the cable and any possible violation of the EMC (electromagnetic compatibility) regulations introduced by the repairing process. When a broken wire is twisted and soldered to repair, it forms a tiny monopole antenna as shown in Fig.~\ref{intro}(b) and (c). This is an inefficient antenna as it is a byproduct of the repair process, not an intentionally designed one. The antenna length is seldom $\frac{\lambda}{4}$. There is no dedicated ground plane with a radius ${>}\frac{\lambda}{2}$; rather, the earth acts as the ground plane for it. Nonetheless, such a monopole antenna helps unintended RF emissions or electromagnetic (EM) emanations transmit to a much longer distance than they otherwise can. This extension in transmission range is a severe threat to data security. The reason is that EM emanations, the byproduct of device operation, contain a significant correlation with the data being processed within devices or carried via cables. Due to this correlation, EM emanations can be processed to extract information, leading to data leakage through an EM side channel \cite{sc0, sc2, sc3}. However, such emissions are inherently weak and often contained within the room where the device or cables are. This work reveals that the repairing process can make them exploitable by a remote attacker who is outside the facility. 

In the conference version of this work \cite{broken_cable}, we intentionally damaged and repaired the three most commonly used cable types (USB, power cable, and display cable) by employing the most commonly used repairing processes: twisting, soldering, and using butt connectors \cite{yt_fix, fix1, butt_con}. We showed that the repaired cables cause significantly louder emanation compared to their normal counterpart, which can be detected even at ${>}$\SI{4}{\meter} distance. Additionally, it is detectable even through a \SI{14}{\centi\meter} thick concrete wall. In this extended version of that work, we have exploited the enhanced emanation due to the repairing process by detecting keystrokes from a repaired (the cable portion only) USB 2.0 keyboard. The reason for specifically choosing modern USB keyboards as the target is that they have trivial EM leakage before any hardware modification. Older PS2 keyboards had a clock wire in them, leading to strong EM leakage and corresponding keystroke detection \cite{ps2_1, ps2_2}. However, the USB protocol excludes the clock signal and uses a differential protocol. Also, the differential wires are twisted together to not only provide immunity against outside interference but also cancel out unintentionally generated emanations of opposite phases. This reduced the leakage from USB cables drastically. Moreover, compared to older low-speed USB cables (USB 1.0), modern faster USB cables (USB 2.0, which can go up to \SI{480}{\mega b p \second}) have a shielding wire and metal shielding wrapper, blocking the emanation signal even further. So, while there are a few publications on keystroke detection from USB keyboards following the USB 1.0 protocol \cite{usb1_1,usb1_2,usb1_3}, there aren't many such publications for USB 2.0 cables. Hence, we selected our target with a strong contrast between the default state (unaltered keyboard) and the modified state (repaired keyboard) in terms of the unintended emission.

We made an extensive dataset by collecting unintended emission data for 70 frequently used keystrokes (0-9, a-z, A-Z, period, comma, space, backspace, CTRL, ALT, SHIFT, and ENTER keys) at different distances from the target in 3 unique environments: (i) open space (no coupling element) (ii) outside an office room (many coupling elements) (iii) outside a building (few coupling elements). We have developed a faster and more efficient detection algorithm for keystroke detection with which we have achieved 100\% keystroke detection accuracy up to \SI{12}{\meter} distance, even through a \SI{14}{\centi\meter} thick wall in between the target and receiver. This is the highest distance-accuracy combination reported in the literature for USB keyboards. Performance variation due to different environments and distances has been studied in detail. Two experimental factors, large interferer and human-body coupling, have been explored, and their impact has been analyzed. Some possible countermeasures have been discussed as well.

\subsection{Our Contribution}
Our specific contributions are as follows:

\begin{itemize}
    \item In the preliminary study, by collecting EM emanation data from three types of repaired cables (USB, power cable, and HDMI), \textbf{we have shown that the traditional repair process creates a monopole antenna as a byproduct, which makes the EM emanation exploitable at a longer range. Although emanation power varies among cables, it can be detected at \bm{${>}4$} m distance for all of them and even through a 14 cm thick concrete wall.} These results show that the repaired cable creates a strong leakage source that can be exploited even from outside of a room.
    \item In this extended version, \textbf{we have formed an extensive dataset by collecting emanation data corresponding to 70 different keystrokes (0-9, a-z, A-Z, period, comma, space, backspace, CTRL, ALT, SHIFT, and ENTER keys) from a USB keyboard in 3 different environments: (i) open space (almost no coupling element), (ii) outside an office room (many coupling elements), and (iii) outside a building (few coupling elements).} An efficient keystroke detection algorithm has been developed to process the captured signals. \textbf{We achieved 100\% keystroke detection accuracy up to \bm{$12$} m distance}, which is the highest reported range with such accuracy for USB keyboards in the literature.
    \item The effect of two crucial experimental factors, \textbf{large interference signals and human-body coupling, has been studied in detail}, and their impact on the overall performance has been analyzed.
    \item As a probable countermeasure to prevent repair-related leakage, \textbf{we have applied an external shielding consisting of 8 layers of aluminum foil. Experimental data show that it reduces emanation power up to \bm{$30$} dB}, but cannot suppress it completely.
\end{itemize}
\subsection{Organization of the Paper}
The rest of the paper is organized as follows: relevant published works are discussed in Section~\ref{sec_rel}. Section~\ref{sec_EI} analyzes the electromagnetic impact of the cable repairing process. It contains all the major results from the preliminary version of this work. This extended version starts from Section~\ref{sec_EXPL}, which discusses our experimental setup, data collection and processing steps for keystroke detection in an open space (on a driveway). It also discusses our detection algorithm and detection ranges. Section~\ref{sec_other_env} explores the keystroke detection process in two different environments: outside an office room and a building. Section~\ref{sec_other_factors} analyzes the impact of two experimental phenomena on keystroke detection performance: large interference and human-body coupling. Section~\ref{sec_counter} discusses possible countermeasures to be included in the repairing process to suppress the enhanced emanation due to the repairing process. Finally, this work is summarized and concluded in Section~\ref{sec_conc}. 

\section{Related Works}
\label{sec_rel}
The leakage property of electromagnetic emanation has been known and exploited by government agencies for a long time. During World War II, a Bell Labs researcher accidentally detected such emission from the Bell 131-B2 mixer. With further investigation, Bell engineers were able to recover 75\% of the typed plaintext by capturing emanations from a Bell 131-B2, placed in a \SI{25}{\meter} away building across the street \cite{ps2_1}. Due to the information leaking property, EM emanation is also known as compromising emanation (CE). There are regulations, called TEMPEST (codename), in both the EU and NATO to protect confidential information against such leakage \cite{nato_tempest, ce_wired_usb}. Its specifications are confidential as well. The first unclassified research paper on emanation was published by Wim van Eck in 1985 \cite{van_eck}, who successfully recovered the screen content of a display unit at a long range using very cheap equipment. Electromagnetic emanations are sometimes called `van Eck radiation' after him \cite{wiki_temp}. In 2002, a detailed investigation on monitor emanations and screen text reconstruction was performed by Markus G. Kuhn in his doctoral dissertation \cite{tech_rep_kuhn}. Since then, a lot of studies have been performed to investigate vulnerabilities associated with EM emanations from various electronic devices.


\begin{figure*}[ht]
\centering
\includegraphics[width=0.98\textwidth]{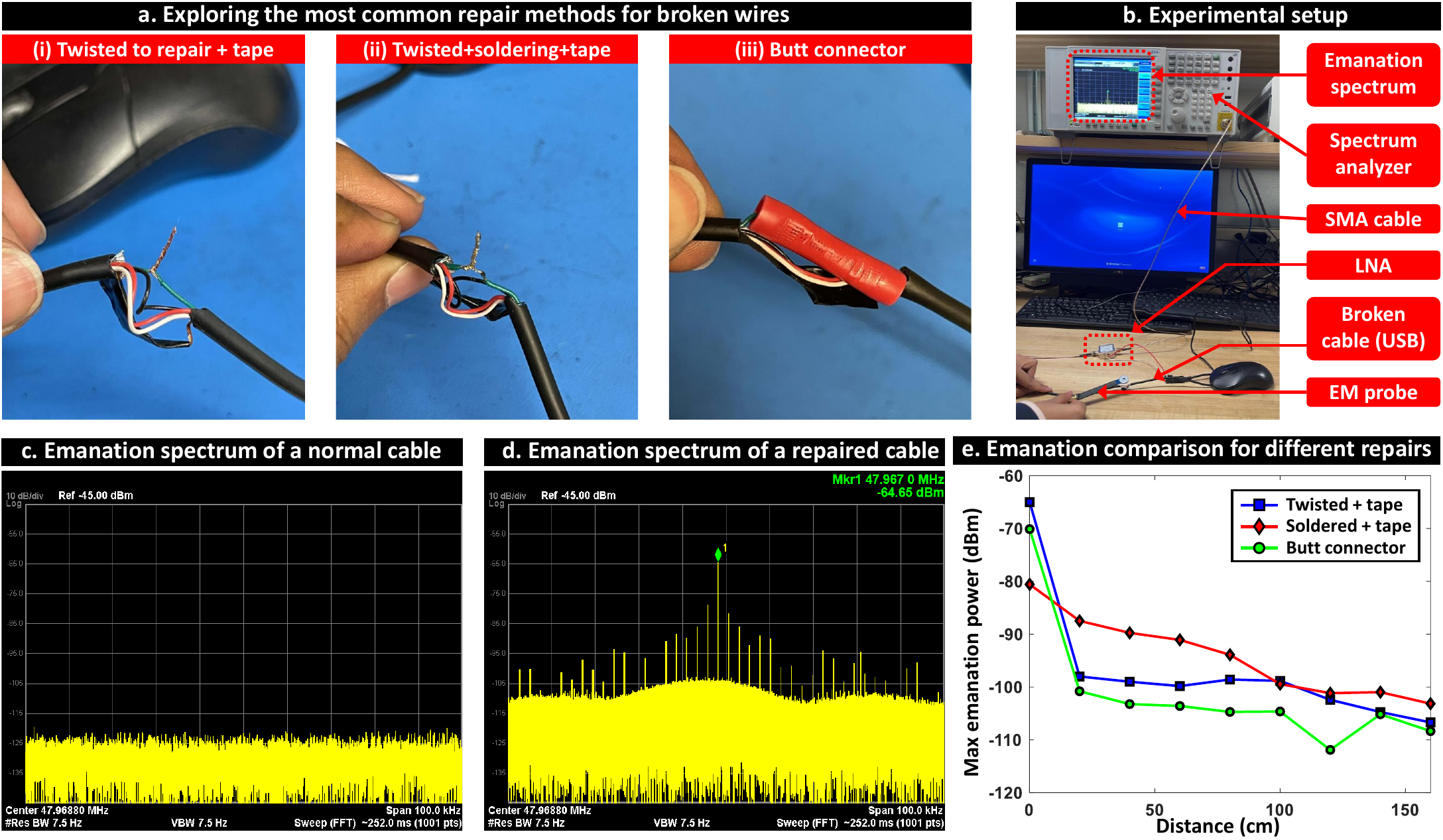}
\caption{(a) Most commonly used repairing methods: twist to join, soldering, and using butt connector. All these methods have been used in this work. (b) The physical experimental setup shows emanation data collection using an EM probe. (c) and (d) Emanation spectrum from normal and repaired USB cables, respectively. It is observed that the normal cable has no significant emanation, while the repaired cable is much louder. (e) Comparison of maximum emanation power from the same USB cable, but repaired using 3 different methods. The plot shows that the soldering method renders a slightly higher SNR emission.}
\label{exp_setup}
\end{figure*}

Emanation from data storage devices has been used to monitor and classify their activity (reading, writing, or silence) \cite{usb_stick}. In \cite{dnn_eman}, authors have exploited GPU emanation to detect DNN architecture. In \cite{smartphone}, emanations from smartphones are used to detect the camera status (both front and rear). Monitor emanation has been exploited to reconstruct a grayscale image containing text and geometric shapes \cite{bari_date23}. In a very recent work \cite{ims2022_eman}, the authors have characterized emanations from an off-the-shelf PC. One interesting phenomenon was frequency shifting with the execution of different programs. This temporal behavior has been exploited in another recent work \cite{arduino_covert} to provide FSK-modulated data for covert communication. There are other studies with more advanced covert communication channels formed using EM emanation \cite{gsmem, airfi, bitjabber, noisehopper}. These studies show the vast attack surface exposed by the EM emanations.

Despite being used mostly for data theft, EM emanation has been used for some useful applications as well. Human-induced EM emanation has been used for access control (touch to unlock) \cite{touch_access}. While traditional RF fingerprinting often involves transmitter-specific physical properties \cite{bari_dirac, bari_rfpuf_ims, bari_rfpuf_fe}, authors in \cite{eman_finger} have used EM emanation to fingerprint IoT devices. Also, they have been used for digital forensics, identification of counterfeit motherboard components, and detection of the presence of a rogue device in a secure facility \cite{eman_forensic, eman_mb, bari_iotj}.

Among many electronic devices, emanation from keyboards got special attention from security researchers since it is the primary input device to the computers, which is used to input sensitive data such as passwords, credit card information, bank account details, and other personal data. Emanations can be of different types, e.g., acoustic, thermal, electromagnetic, etc. We have heard the typing sounds from keyboards. As it turns out, each keystroke produces a slightly different sound signature, forming an acoustic side-channel. Researchers have exploited this acoustic emanation for keystroke recovery \cite{akb1, akb2, akb3, akb4, akb5}. Thermal residue on keys has been exploited to detect typed passwords even after \SI{30}{\second} from initial typing \cite{thermal_kb}. Authors in \cite{kb_motion} have developed a motion-based keystroke recovery method that can recover 4-digit PINs with 65\% probability out of 81 tries. Researchers have proposed a WiFi-signal-based keystroke recovery method called WiKey, which achieves 96.4\% accuracy in single-key detection \cite{kb_wifi}. Human-coupled EM emanation from touchscreens has been exploited to infer user key (PIN) input \cite{periscope}. This method, called Periscope, recovered 6-digit PINs with 56.2\% accuracy.

Other works on keystroke recovery focus on exploiting EM emanation or compromising emanation specifically. Older PS2 keyboards had a separate clock signal carrying wire, leading to significant EM emanation. Using that, authors in \cite{ps2_1} recovered keystrokes with 95\% accuracy in an office environment and a residential building. They proposed four different detection methods by studying PS2, wireless, USB, and laptop keyboards. However, their method, while successful for PS2 and laptop keyboards, could not detect the individual keys for USB or wireless keyboards. Instead, the proposed method reduced the probable key space only (${\sim}$5 probable keys on average per detection). A similar method was used in \cite{ps2_2}, where the authors detected keystrokes from a PS2 keyboard at \SI{1.5}{\meter} distance. However, an interesting conclusion from this work was that even shielding couldn't suppress the compromising emanations completely. Authors in \cite{usb1_1} have developed an automatic keystroke detection method from EM emanation using the autocorrelation function. Their optimistic (\SI{0}{\deci\bel} building attenuation) theoretical calculation showed that EM emanation from USB keyboards can be detected at \SI{11}{\meter} range in free space. Authors in \cite{usb1_2} have collected emanation using an H-field probe and an oscilloscope. They applied peak detection and transformed the collected peaks into a bit pattern to recover keystrokes. Researchers in \cite{usb1_3} have detected 32 out of 36 target keystrokes (alphanumeric keys only) at \SI{0.15}{\meter} away from the target USB keyboard using EM emanation. There are a couple of analytical works that involved the development of an automatic control system for emanation reception \cite{th_kb_1} and the significance of a priori information for the quality of detected emanation \cite{th_kb_2}.

\section{Electromagnetic Impact of Repairing Cables}
\label{sec_EI}
\subsection{Broken Cable Repairing Methods}
\label{repair_method}
\begin{figure*}[ht]
\centering
\includegraphics[width=0.98\textwidth]{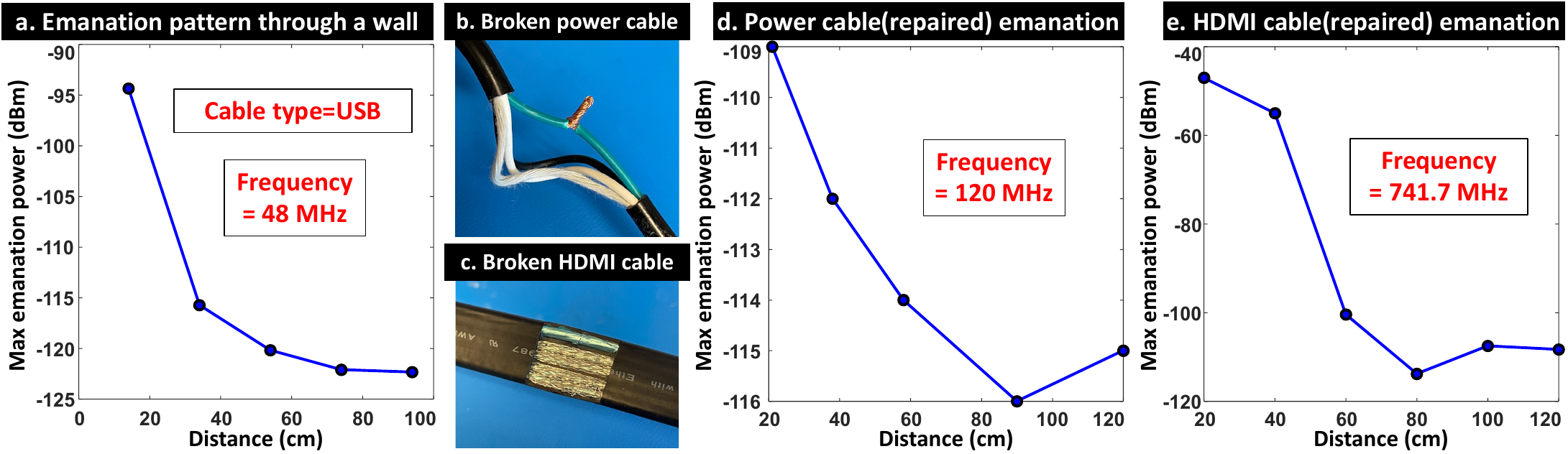}
\caption{(a) Emanation pattern of a USB cable detected through a \SI{14}{\centi\meter} thick concrete wall. (b) A broken power cable has been twisted to repair. (c) A repaired HDMI cable with metal shielding removed. (d) Emanation pattern of a repaired power cable. (e) Emanation pattern of a repaired HDMI cable.}
\label{wall_hdmi}
\end{figure*}
There are many different methods to fix the broken wires inside a cable \cite{yt_fix, fix1}. The most common method involves scraping off a portion of insulation from the severed wire to expose the metal strings and twist them together as shown in Fig.~\ref{exp_setup}(a)(i). After joining them, electrical tape or a heat shrink tube is used to restore insulation. Fig.~\ref{exp_setup}(a)(ii) shows a modified version of this method where the broken pieces of the wire are soldered together before applying insulation. Another method is to use a butt connector \cite{butt_con} which is a plastic tube with a hollow metal tube inside. The broken pieces of wires are inserted in it from both sides and then the connector is pressed hard so that the internal metal tube gets squeezed and holds the wires firmly together. Fig.~\ref{exp_setup}(a)(iii) shows the use of a butt connector. In this work, we have applied all these repair methods.

\subsection{Data Collection}
Fig.~\ref{exp_setup}(b) shows our experimental setup in our lab. An EM probe is used as a sensing device to collect emanation data from the target (broken cable). The EM probe is connected to a 32 dB low-noise amplifier (LNA) which amplifies the signal and sends it to a spectrum analyzer where the signal is collected and analyzed. We have intentionally broken different types of cables by stripping off their external plastic layer and cutting either the data wire or the power-carrying wire. Then the cables are repaired using different methods described above in subsection~\ref{repair_method}. Fig.~\ref{exp_setup}(c) and (d) show the emanation spectrum from the normal and repaired USB cable, respectively.

\subsection{Emanation from USB Cable}
To test emanation from USB cables, a USB mouse was used. It was USB 2.0, which supports a 12 Mbps data rate at full speed. Hence, 12 MHz and its harmonics were our primary targets. The strongest emanation was found at the $4^{th}$ harmonic (48 MHz), which is shown in Fig.~\ref{exp_setup}(d). Data were collected from the nearest position to the cable up to 4 m, in 20 cm steps. Fig.~\ref{exp_setup}(e) shows the maximum emanation power versus distance plot for up to 160 cm for USB cables with three types of repairing process (twisting, soldering, and using butt connector). The soldered wire seems to have a slightly stronger emanation compared to the other two. This is not surprising since soldering forms a more unified antenna. 

\subsection{Detection through Obstacles}
Next, the detection is performed through a wall to imitate an eavesdropper outside of the facility. For that purpose, the broken cable is kept outside of the experiment room while the emanation signal is collected from inside. Fig.~\ref{wall_hdmi}(a) shows collected emanation power through the wall versus distance (includes wall thickness). The wall causes ${\sim}$20 to 30 dB loss. However, it is still detectable, meaning sensitive data can be stolen even from outside the room.

\subsection{Emanation from Other Cables}
\subsubsection{Power Cables}
After exploring USB cables, we focus on other cables and check if the same phenomenon occurs for them. Fig.~\ref{wall_hdmi}(b) shows a repaired power cable. The cable was supplying power to a desktop to which the mouse was connected. The desktop contains an intel\textregistered\ core$^{\text{TM}}$ i7-6700 microprocessor and \SI{8}{\giga\byte} RAM. For this power cable, we found the strongest emanation at \SI{120}{\mega\hertz}. Fig.~\ref{wall_hdmi}(d) shows the maximum emanation power vs distance for the broken power cable. Here, the emanation follows a similar pattern as in the case of the USB cable, but weaker.
\subsubsection{HDMI Cables}
Next, emanations from a display cable (HDMI) are tested. Fig.~\ref{wall_hdmi}(c) shows a broken HDMI cable with internal shielding removed. Emanation from the HDMI cable is the strongest. Unlike the other two cables, HDMI has detectable emanation without breaking the cable. However, the repairing process makes the emanation even stronger. On top of the cable, the strongest detected emanation was ${\sim}$-20 dBm. Even at 20 cm, it was -47 dBm. Fig.~\ref{wall_hdmi}(e) shows the maximum emanation power vs distance. As mentioned in section~\ref{sec_rel}, emanations from regular HDMI cables and monitors have been exploited to reconstruct screen images and texts. Repairing methods make the screen content recovery easier with higher emanation SNR.

\section{Exploiting Emanation from Repaired Cables - Keystroke Detection}
\label{sec_EXPL}
\subsection{Data Transfer Protocol for USB Keyboard}
Most widely used USB keyboards follow the USB 2.0 protocol. This protocol has 3 speeds that it can run at: (i) low speed (1.5 Mbps), (ii) full speed (12 Mbps), and (iii) high speed (480 Mbps). Most standard USB keyboards transfer data at full speed, while some gaming keyboards use high speed. For our experiment, we have used a Dell KB216 QWERTY keyboard, which follows the USB 2.0 protocol and runs at full speed. Unlike older PS2 keyboards, USB keyboards don't have a separate clock signal. Rather, it uses a differential signal protocol which is NRZI encoded, meaning a change in the differential signal represents bit `0' and no change represents bit `1'. Differential signal provides a much higher immunity against external interference. Also, it cancels unintentionally generated electromagnetic emanation as they are \SI{180}{\degree} out of phase with each other. Another interesting feature is `Bit Stuffing'. Since there is no clock signal for synchronization, too many consecutive bits of the same value may cause bit errors. To avoid that, A `0' bit is inserted after every 6 consecutive `1' bits in the data stream.

\begin{figure*}[ht]
\centering
\includegraphics[width=0.98\textwidth]{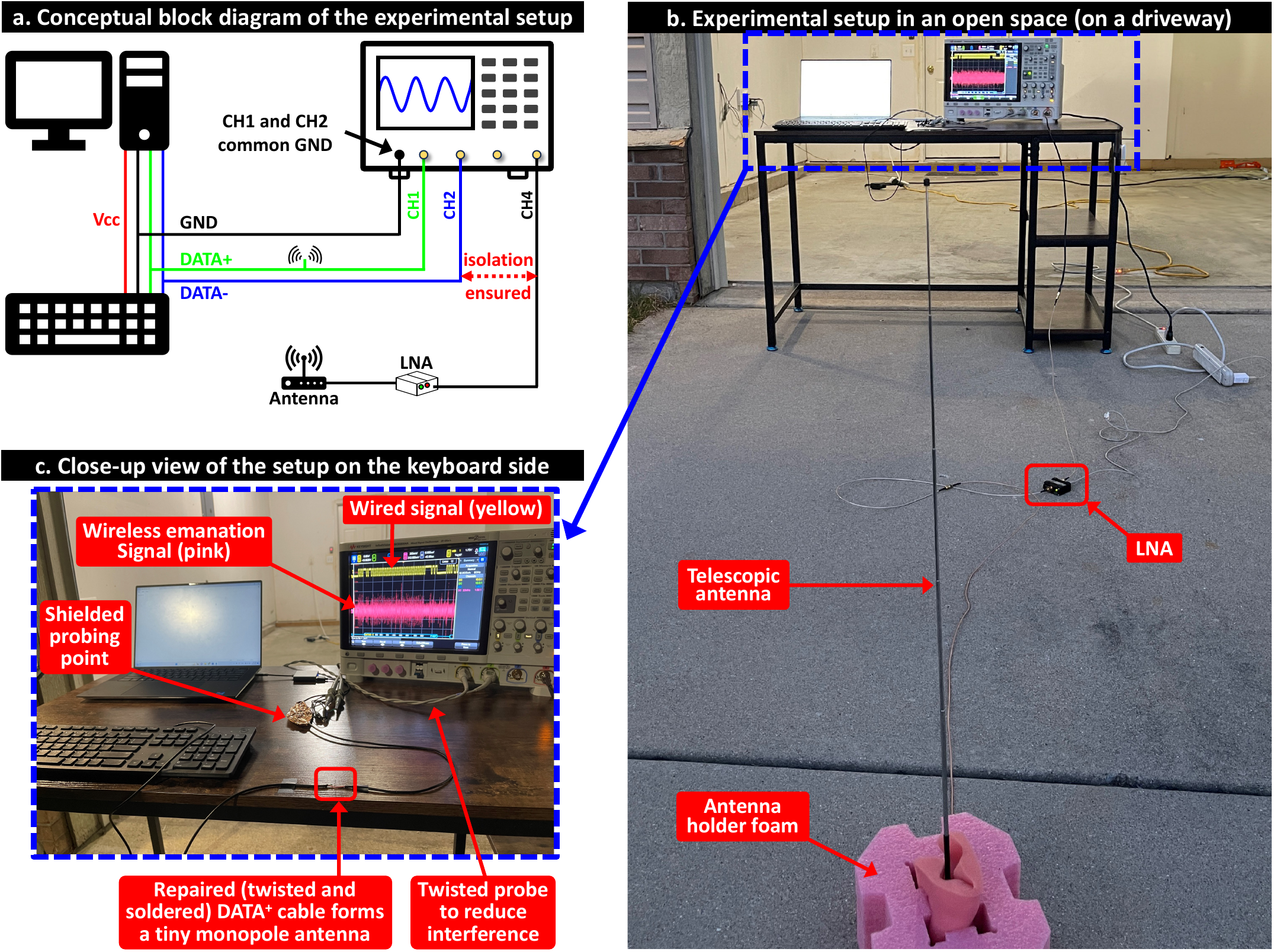}
\caption{(a) Conceptual block diagram of our experimental setup to collect data. (b) Actual setup in an open space (on a driveway) with almost no coupling element nearby (c) A close-up view of the setup showing the tiny monopole antenna formed due to the repairing process. It also shows the shielding at the probing point and both probed signal and wireless emanation signals on the oscilloscope.}
\label{drv_exp}
\end{figure*}

For communication, the keyboard remains in an idle state, and the computer scans for any keystroke at a regular interval (usually \SI{16}{\milli\second} for low speed, \SI{1}{\milli\second} for high speed). There are usually 4 types of data packets:
\begin{itemize}
    \item IN Packet: the computer asks for input
    \item DATA Packet: contains keystroke data from the keyboard to the computer
    \item Handshake Packet: acknowledgment of data
    \item NAK: basically, the keyboard says no change in keystroke
\end{itemize}
Each packet starts with a SYNC pattern and ends with a CRC. Depending on the packet type, there might be an input address, actual keystroke data, acknowledgment, or error code in between.

At the hardware level, USB 2.0 also uses a 4-wire system, which is color-coded as follows:
\begin{itemize}
    \item VCC: Red wire
    \item GND: Black wire
    \item DATA+ or D+: Green wire
    \item DATA- or D-: White wire
\end{itemize}
There is also an uncoated shield wire that is connected to the metal shielding foil wrapping around these internal wires. The whole wire system is put inside an insulating PVC jacket. The metal shielding and differential protocol lead to trivial electromagnetic emanation from USB keyboards, unlike the older PS2 keyboards, which had very high leakage. The documentation of the whole USB 2.0 protocol can be found in \cite{usb2protocol}
\subsection{Experimental Setup}
Fig.~\ref{drv_exp}(a) shows the conceptual block diagram of our experimental setup. The DATA+ wire (green wire) is intentionally cut and repaired (by soldering) to form a tiny monopole antenna, which causes the generated electromagnetic emanation to radiate better. An antenna collects the radiated emission, amplifies the collected signal using a low-noise amplifier (LNA), and sends it to channel 4 of an Oscilloscope, where it is captured and stored. However, as mentioned earlier, the computer scans for any keystroke at every \SI{1}{\milli\second} interval (default polling rate may vary slightly depending on the keyboard model and the operating system). Within this period, the actual frame (consisting of several packets) is only \SI{80}{\micro\second} where each bit is \SI{\sim0.08}{\micro\second}. Hence, triggering the oscilloscope at the scanning signal doesn't provide details of the bit patterns within the frame. Triggering the oscilloscope at the frame edge is also infeasible, as most of the time the signals are in an idle state (differential `1').

Our goal in this work is to demonstrate how we can exploit the enhanced emanation from a repaired keyboard instead of building a perfect receiver specialized for this type of data collection. Hence, two probes from channels 1 and 2 of the oscilloscope (they have a common ground shorted internally within the oscilloscope) are connected to the data wires (DATA+ and DATA-) and GND. The scope then generates a trigger signal at the handshake packet of the whole frame. This method or a similar one has been used in published literature \cite{ps2_1, usb1_2}, and we followed the same setup. The question arises as to whether the wired connection somehow contaminates the wireless data. Several steps have been taken to ensure the isolation between the wired and wireless data within the scope:
\begin{itemize}
    \item Wired data is connected to CH1 and CH2 (which have common ground), while wireless data is captured at CH4, which has a separate ground. Since the keyboard ground is provided by the computer, we used a laptop (during data collection, it was disconnected from the AC supply and was powered by its battery).
    \item The probing points were insulated, and the whole portion was shielded by multiple layers of copper foil. The shielding was connected to the shielding wire. This suppressed the radiation from the probing points. The CH1 and CH2 probes were also twisted to cancel out most interferences. 
    \item When the antenna was disconnected from the LNA, CH4 (channel 4) of the oscilloscope immediately turned to a noisy flat line, confirming that the signal it was showing was indeed what the antenna was collecting.
\end{itemize}

Fig.~\ref{drv_exp}(b) shows the actual setup in a driveway that has open space on both sides. Why a driveway instead of a lab? Our initial testing showed that the radiated emission can couple with nearby devices/elements and travel over long distances. An open space with no such coupling element nearby ensures that we can detect its worst-case transmission range. In section~\ref{sec_other_env}, we explored other environments (outside an office room and a building, respectively) which are more likely in the practical attack scenarios. A telescopic antenna has been held by a foam box, which captures the radiated emission. The EM emanation signal is amplified using an LNA before being sent to the scope.

Fig.~\ref{drv_exp}(c) shows a close-up view of the scope side. It shows the tiny monopole antenna formed by the repaired wire (DATA+). It also shows the twisted oscilloscope probes and the shielded probing point. The wire data (only DATA+ shown) is shown in yellow, whereas the wireless emanation signal is shown in pink on the oscilloscope.

\begin{figure*}[htbp]
\centering
\includegraphics[width=0.98\textwidth]{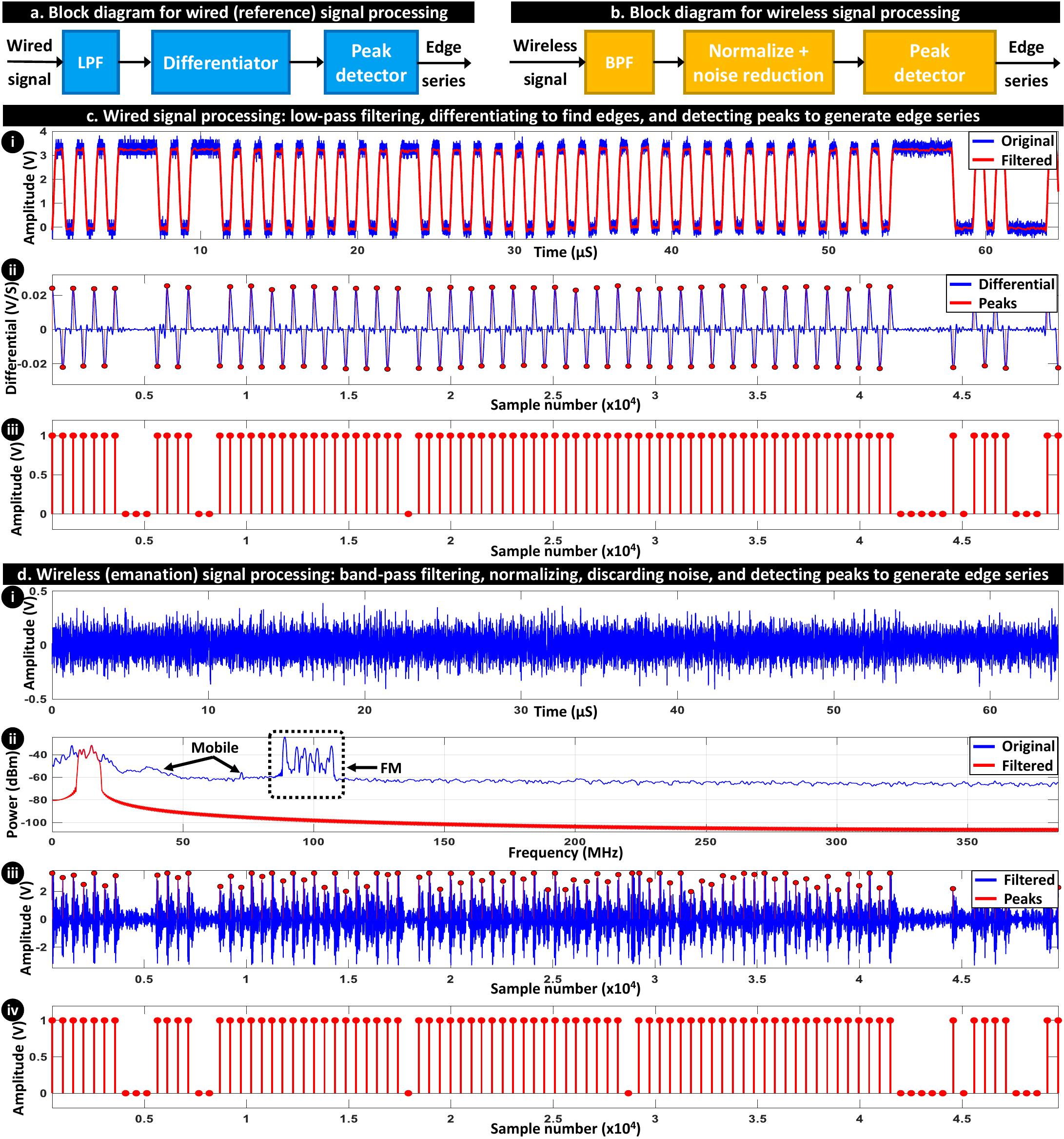}
\caption{Block diagram showing signal processing steps for (a) wired and (b) wireless data for keystroke `a'. (c)(i) shows the DATA+ wire signal (blue) and the low-pass filtered version of it (red). The filtered signal is differentiated to detect the rising and falling edges. A peak detection algorithm then provides the location of the edges as shown in (c)(ii). Finally, a binary series is formed where `1' represents the presence of the edge and `0' represents its absence. (c)(iii) shows the edge series. The edge series for all 70 keys has been derived once and saved as references. (d)(i) shows the collected emanation data. It looks mostly noisy. However, spectrum analysis in (d)(ii) reveals that there are a lot of interferences from FM radio, mobile signal, and other sources. These are filtered out using a bandpass filter. Since our data rate is 12 Mbps, the emanation signal is expected to be around \SI{12}{\mega\hertz}. So, the cut-offs are set at \SI{10}{\mega\hertz} and \SI{18}{\mega\hertz}. The filtered signal is then normalized to be within $A=\pm3.3$V. (d)(iii) shows the filtered signal (blue), which looks much cleaner. The signal is further cleaned by setting any amplitude lower than $\frac{A}{2}$ to 0. Finally, peak detection is performed to detect the edges. (d)(iv) An edge series is formed from the detected edges. By comparing with the reference edge series at (c)(iii), we can see that they match perfectly.}
\label{data_process}
\end{figure*}

\begin{figure*}[ht]
\centering
\includegraphics[width=0.98\textwidth]{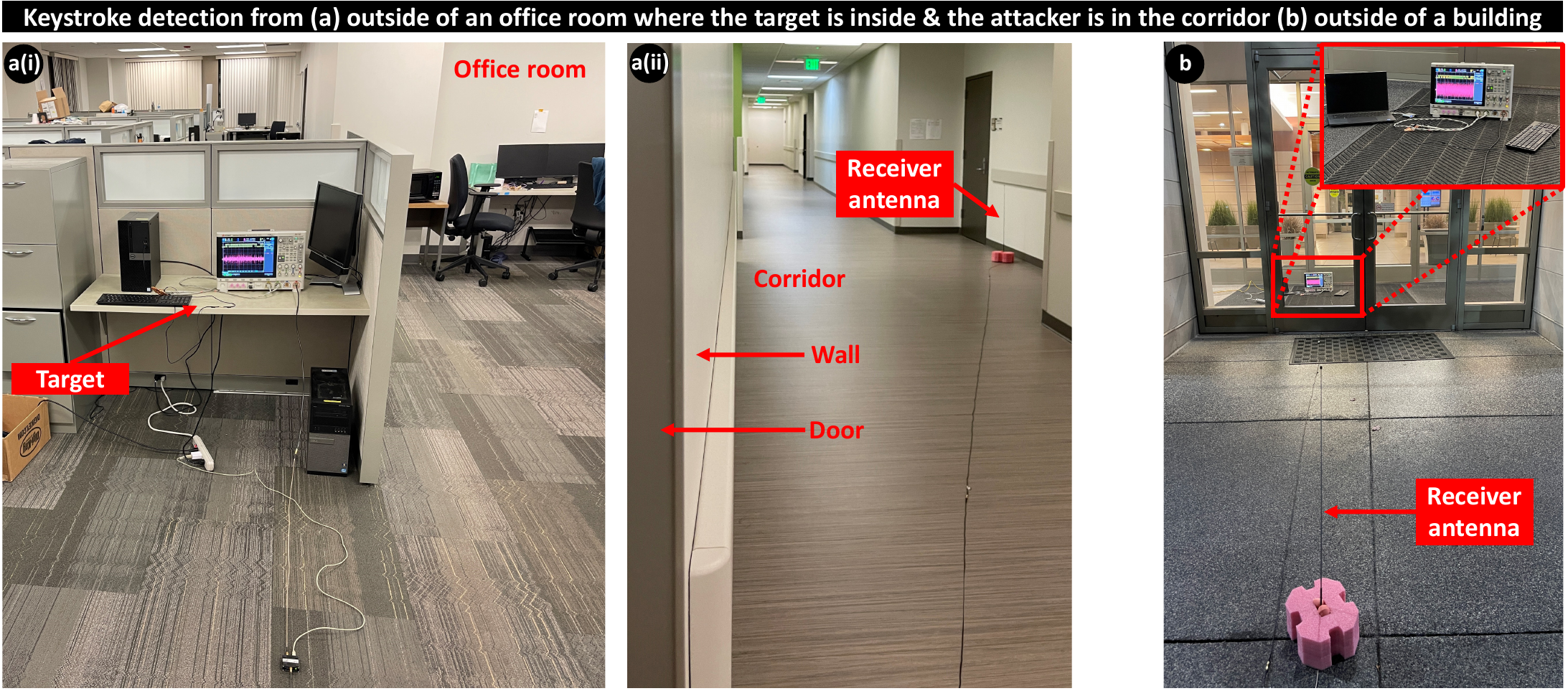}
\caption{(a) Experimental setup for data collection from outside of an office room. The target is kept inside the room (i) and the receiver outside (in the corridor) (ii), while there is a \SI{14}{\centi\meter} thick wall in between them. (b) Experimental setup for data collection from outside a building. The target is kept inside the building while the receiver is outside.}
\label{other_setup}
\end{figure*}

\subsection{Data Collection}
Modern keyboards usually have 104 keys, although that number varies slightly among different keyboard models. To keep the data collection time and dataset size reasonable while covering the most commonly used keys, we have collected data for 70 different keystrokes. Our proposed algorithm is scalable as it detects the original bit pattern. Hence, it can be used for any number of keys. The 70 keys are as follows:
\begin{itemize}
    \item 10 numeric keys: from 0 to 9
    \item 52 alphabetic keys: 26 letters in both uppercase and lowercase formats.
    \item 8 most commonly used special keys: Period, Comma, Space, Backspace, CTRL, ALT, SHIFT, and ENTER
\end{itemize}

Emanation data for the whole keystroke set is collected at least twice at each distance, for 4 different distances from the target: \SI{0.5}{\meter}, \SI{2.5}{\meter}, \SI{3}{\meter}, and \SI{3.8}{\meter}. Later, the collected data are processed in MATLAB.

\subsection{Data Processing}
Fig.~\ref{data_process}(a) and (b) show the signal processing steps in separate block diagrams for wired and wireless signals, respectively. Fig.~\ref{data_process}(c) shows the wired signal (collected from the DATA+ wire) processing steps for keystroke `a'. At first, some samples from the beginning and end are discarded, which are just flat lines with no bit pattern. Then the signal is filtered using a lowpass filter of \SI{5}{\mega\hertz} cut-off. The blue and red plots in Fig.~\ref{data_process}(c)(i) show the probed data and filtered version of it, respectively. Since emanation is generated due to signal switching at the rising and falling edges, we need to find the location of those edges. To get that information, the filtered signal is differentiated. Fig.~\ref{data_process}(c)(ii) shows the differentiated signal (blue plot). Then, peak detection is performed to find the peaks corresponding to edges. Finally, an edge series is generated, which is essentially binary data about edges. At each bit interval, if there is an edge, the series is assigned a value of `1', else it is assigned `0'. Fig.~\ref{data_process}(c)(iii) shows the edge series. This series is generated only once for all 70 keys and is saved as a reference.

Next, wireless emanation signals are processed. Fig.~\ref{data_process}(d) shows the processing steps in detail. The collected wireless signal is plotted in Fig.~\ref{data_process}(d)(i). At first glance, it appears to be mostly noise. Since the time-domain signal doesn't reveal much information, we switched to the frequency domain and looked at the power spectrum. Fig.~\ref{data_process}(d)(ii) shows the power spectrum of the collected signal in the blue plot. The plot reveals that there is a strong interference of FM radio signal in the \SI{88}{\mega\hertz} to \SI{108}{\mega\hertz} range. Also, there is interference from mobile signals as well. We already know that our USB keyboard runs at \SI{12}{\mega b p\second}. We expect the emanation to be in that range. Hence, the signal is filtered using a bandpass filter with lower cut-off=\SI{10}{\mega\hertz} and upper cut-off=\SI{18}{\mega\hertz}. The red plot in Fig.~\ref{data_process}(d)(ii) shows the filtered spectrum. 

Next, the filtered signal is normalized to $A=\pm3.3V$. Detecting the normalization factor was a bit tricky. In the collected data, we sometimes observed a strong spike/glitch from an unknown interference source (this phenomenon is discussed in detail in section~\ref{subsec_inter}). Hence, we took the maximum signal amplitude, $S_{max}$ from the lower 99\% data (skipping the top 1\% value so that the strong interference spike doesn't affect normalization). The normalizing factor is determined as $\frac{A}{S_{max}}$. Any signal value above $A=3.3V$ is clipped to $A$. Fig.~\ref{data_process}(d)(iii) shows the filtered and normalized signal in the time domain (blue plot). Also, any value lower than $\frac{A}{2}$ is set to zero (to discard the noisy values) before peak detection. The red stem plot in Fig.~\ref{data_process}(d)(iii) shows the detected peaks (which also indicates the detected edges). From the peaks, the edge series is formed. At each reference edge location, if there is a peak detected in proximity to that location (they don't appear at the perfect location most of the time, rather slightly shifted left or right), a value of `1' is assigned. If not, a `0' is assigned to the series. Fig.~\ref{data_process}(d)(iv) shows the edge series generated from the emanation signal. A comparison between Fig.~\ref{data_process}(a)(iii) and Fig.~\ref{data_process}(d)(iv) reveals that they are identical.

\subsection{Detection Algorithm}
The computer scans the keyboard at every \SI{1}{\milli\second} interval. Hence, a real-time keystroke detector must be efficient to render high accuracy, yet fast enough to perform the detection steps within this time frame. Our detection algorithm has two parts. In the first part, the wired signal is differentiated and the reference edge series is generated for all 70 keys. This is done only once. The references are saved. In the second part, the collected data are passed through a bandpass filter. It can be an active filter at the receiver front so that the captured data already comes as filtered, reducing the processing burden. All the DSP processor has to do is detect the peaks, form an edge series, and compare them against the 70 references. The reference series that shows the highest match is the detected key. This method is very fast with less computational requirement compared to many spectrogram-based methods suggested in some other works. 

\begin{figure*}[ht]
\centering
\includegraphics[width=0.98\textwidth]{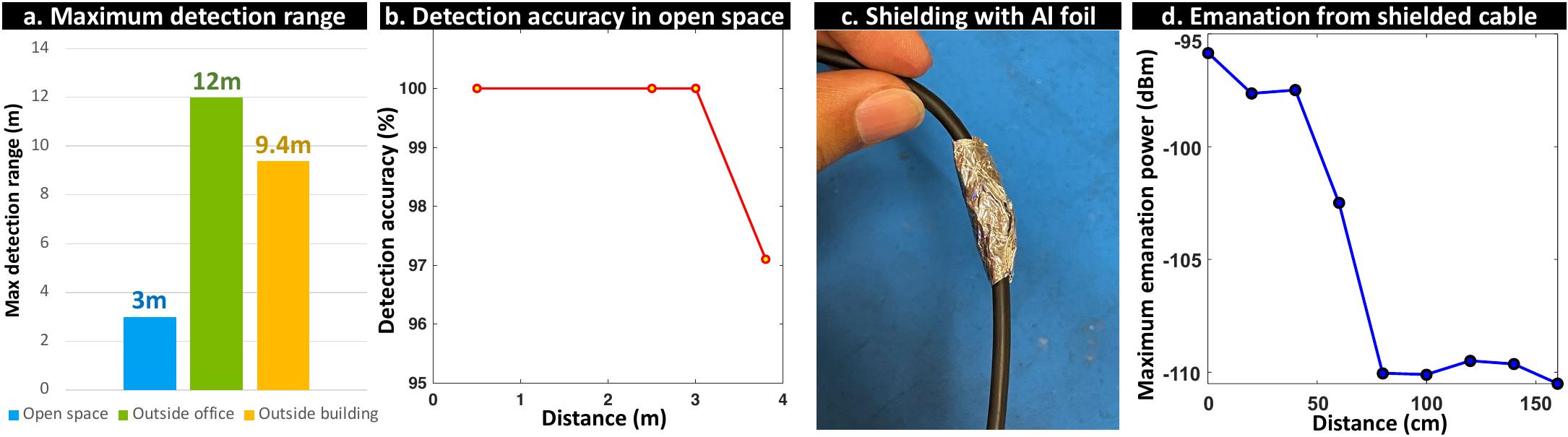}
\caption{(a) Maximum detection range (up to the range of ${\sim}100$\% accuracy) in 3 different environments: open space, outside the office, and outside the building. (b) Variation of keystroke detection accuracy with distance in an open space environment. (c) A `broken and twisted to repair' USB cable is externally shielded with 8 layers of aluminum foil. (d) The emanation pattern of the shielded cable shows reduced emission power.}
\label{shield}
\end{figure*}

\subsection{Detection Range}
Data were collected at \SI{0.5}{\meter}, \SI{2.5}{\meter}, \SI{3}{\meter}, and \SI{3.8}{\meter}. Up to \SI{3}{\meter}, we consistently achieved 100\% accuracy with all the keys being detected accurately. At \SI{3.8}{\meter}, the correct number of detected keys varies from 68 to 70 (97.1\% accuracy). If 2 letters in a paragraph are replaced with 2 other letters, the text is still intelligible. However, if it is a password, even one wrong letter is unsatisfactory. So, the acceptance of this accuracy level depends on the specific application. For most cases, we take \SI{3.8}{\meter} as the longest detection range for keystroke detection in an open space.

\section{Keystroke Detection in Other Environments}
\label{sec_other_env}
\subsection{Keystroke Detection from Outside of an Office Room}
Fig.~\ref{other_setup}(a) shows our experimental setup where the target (monopole antenna from the repaired keyboard), along with the oscilloscope, is kept inside an office room and the receiver antenna is kept in the outside corridor. There are a lot of electronic devices within the office. The electromagnetic emanation wave can couple with them and travel further. Also, the corridor acts somewhat like a waveguide, helping the wave travel a bit better. Considering these factors, we expect to see a longer detection range for this environment. Experimental data shows that we can detect the keystrokes with 100\% accuracy at \SI{12}{\meter} linear distance. A special note here is that there is a $\sim$\SI{14}{\centi\meter} thick wall between the transmitter and receiver here that causes path loss. In a direct line of sight, the range can be even higher.

\begin{figure*}[ht]
\centering
\includegraphics[width=0.98\textwidth]{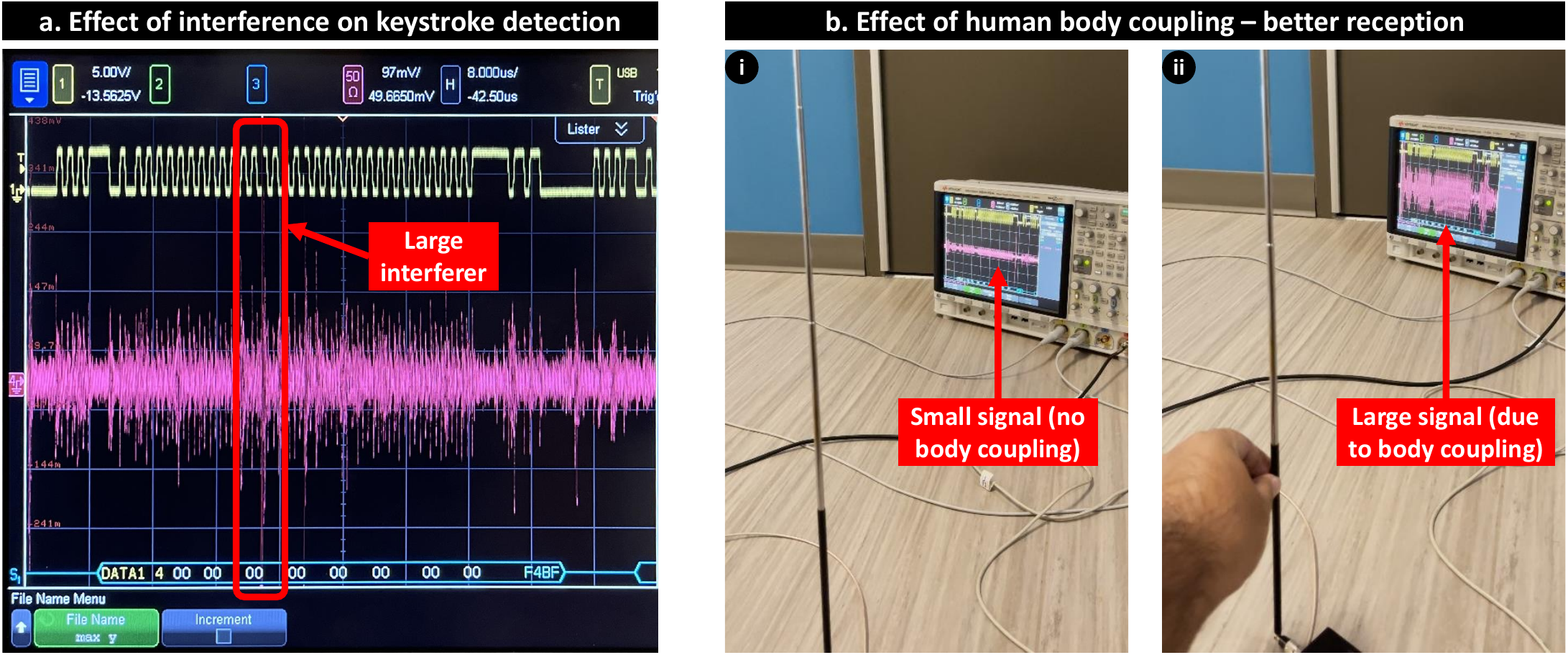}
\caption{(a) Large interference spike/glitch. Depending on its location and number, such interference might lead to a few wrong key detections. (b) Human-body coupling provides better reception, which the attacker may exploit if he/she has continued physical access to the receiver. }
\label{exp_factors}
\end{figure*}

\subsection{Keystroke Detection from Outside of a Building}
Fig.~\ref{other_setup}(b) shows our setup for keystroke detection from the outside of a building. While there are a few coupling elements inside the building, there is none outside. So, this kind of environment is somewhat in between the previous two environments. Our collected data show that we can detect up to \SI{9.4}{\meter} ($31{ }ft$ to be precise) reliably. False detection starts to happen after that range.

\subsection{Detection Range Comparison}
Fig.~\ref{shield}(a) shows our maximum detection range (considered up to the range where keystroke detection accuracy is ${\sim}$100\%) and Fig.~\ref{shield}(b) shows the accuracy versus distance plot for open space environment. In this subsection, our work is compared against published literature in terms of the maximum distance of keystroke recovery from USB keyboards in an office environment. Table~\ref{table_1} shows the comparison. Thanks to the impact of the repairing process, which helps transmit EM emanation much further than otherwise possible.

\begin{table}[h!]
\centering
\caption{Keystroke recovery range comparison}
\begin{tabular}{||c | c ||} 
 \hline
 Detection method & Max range (m) \\
 \hline
 Matrix scan method \cite{ps2_1} & 3 \\
 \hline
 Autocorrelation method \cite{usb1_1} & 1\\
 \hline
 Peak detection and binary & 0.15\\
 code transformation \cite{usb1_3} & \\
 \hline
 This work & 12 \\
 \hline
\end{tabular}
\label{table_1}
\end{table}

\section{Experimental Factors Affecting Performance}
\label{sec_other_factors}
\subsection{Interference}
\label{subsec_inter}
Fig.~\ref{exp_factors}(a) shows an unusually large glitch in the collected emanation data. The source of this large interference remained undetected. However, this glitch can have an impact on the detection. In the best case, it will overlap with another edge, and no harm is done as it is detected as part of that edge. However, it gets tricky if it is between the edges. During the normalization step after filtering, the signal is capped at $A=\pm3.3V$. The peak detector might get confused with this maximum value in between two edges and detect it as another peak. To avoid it as much as possible, the minimum separation between two peaks was specified as $\frac{2}{3}$ of the bit width. However, it gets trickier if such a glitch appears where there are no edges. In the worst case, there might be several of these glitches at such locations. This leads to erroneous `1'(s) in the edge series. Since the edge series has 95-97 edges, the detector often detects the correct keystroke even in the presence of 1 edge mismatch. However, as the number of mismatches increases, the detector can make false detections. 

\subsection{Effect of Human Body Coupling}
Fig.~\ref{exp_factors}(b) shows an interesting phenomenon. Whenever a human touches the non-conductive parts of the receiver antenna, the received signal amplitude increases by $2\times$ to $4\times$. The emanation signal couples with the human body to provide better reception. Also, it acts as an RC filter to suppress some high-frequency noise without any data processing. Attackers may exploit this phenomenon if they can physically maintain continued access to the receiver. 

\section{Countermeasures}
\label{sec_counter}
Shielding is commonly used inside a cable to reduce the emission power. But how effective is external shielding on a broken cable? To answer that, an external shielding is attached to the top of a repaired USB cable. This shielding contains eight layers of aluminum foil. Fig.~\ref{shield}(c) shows the shielded cable. Next, emanations from the cable are measured following our earlier method. Fig.~\ref{shield}(d) shows the maximum emanation power over distance. A comparison of Fig.~\ref{exp_setup}(e) and Fig.~\ref{shield}(d) shows that shielding decays emanation to some extent (up to \SI{30}{\deci\bel}), but can not suppress it totally. However, this suppression reduces the maximum detection range and renders eavesdropping or exploitation of EM emanations harder. So, traditional repairing methods can be augmented with an extra layer of shielding and insulation (e.g., heat shrink tube) for better protection against EM leakage.

As an extra precaution, electronic equipment and the corresponding cables should be placed toward the center of the room. This effectively sets a control perimeter within which the emanation signal decays significantly. However, for an extremely sensitive facility, the best approach is to replace the damaged cable.

\section{Conclusion}
\label{sec_conc}
In this work, we have explored the electromagnetic aspects of traditional repair methods for broken cables. We postulated that the repairing process creates a monopole antenna, which will increase the SNR of the electromagnetic emanation. Employing three types of most commonly used repair methods (twisting, soldering, and butt connector) on different cables (USB, power cable, and HDMI), we have proved our hypothesis experimentally. Our results show that for all types of cables, EM emanation is detectable at ${>}4$ m distance. It is so strong that it can penetrate a concrete wall and remains detectable through a \SI{14}{\centi\meter} thick concrete wall. The strengthened emanation has been exploited for information leakage in the form of keystroke detection. Emanation data from a repaired (the cable portion) USB keyboard have been collected for 70 different keystrokes (0-9, a-z, A-Z, period, comma, space, backspace, CTRL, ALT, SHIFT, and ENTER keys) at different distances and three unique environments: (i) an open space (ii) outside of an office room (iii) outside of a building. A computationally efficient, real-time detection algorithm has been developed that provides 100\% keystroke detection accuracy up to \SI{12}{\meter} distance. This is the highest reported range with such high accuracy for keystroke detection using a USB keyboard. Experimental factors, e.g., interference, human-body coupling, and their impact on detection performance have been analyzed. We have explored a possible remedy, external shielding using metal foil, which suppresses the emanation to some extent but cannot block it completely. This work exposes a previously unexplored security vulnerability that may go unnoticed and affect an otherwise secure cyber-physical system.

\section*{Acknowledgments}
This research was supported by the Office of the Director of National Intelligence (ODNI), Intelligence Advanced Research Projects Activity (IARPA), via contract: 2021-21062400006. The views and conclusions contained herein are those of the authors and should not be interpreted as necessarily representing the official policies or endorsements, either expressed or implied, of the ODNI, IARPA, or the U.S. Government. The U.S. Government is authorized to reproduce and distribute reprints for Governmental purposes not withstanding any copyright annotation thereon.





\begin{thebibliography}{00}


\bibitem{broken_cable}
M. F. Bari, M. R. Chowdhury and S. Sen, ``Is Broken Cable Breaking Your Security?," 2023 IEEE International Symposium on Circuits and Systems (ISCAS), Monterey, CA, USA, 2023, pp. 1-5, doi: 10.1109/ISCAS46773.2023.10181751.
\bibitem{scisrs}
IARPA, ``SCISRS: Securing Compartmented Information with Smart Radio Systems," Iarpa.gov, 2019. https://www.iarpa.gov/research-programs/scisrs (accessed Aug. 23, 2025).
\bibitem{sc0}
G. Camurati, S. Poeplau, M. Muench, T. Hayes, and A. Francillon, ``Screaming Channels: When Electromagnetic Side Channels Meet Radio Transceivers,” in Proceedings of the 2018 ACM SIGSAC Conference on Computer and Communications Security, 2018, pp. 163–177, doi: 10.1145/3243734.3243802.
\bibitem{sc2}
J. Danial et al., ``EM-X-DL: Efficient Cross-Device Deep Learning Side-Channel Attack with Noisy EM Signatures,” ACM Journal on Emerging Technologies in Computing Systems, 2020, vol. 18, no. 1, pp. 1–17. doi: 10.1145/3465380.
\bibitem{sc3}
D. Das, A. Golder, J. Danial, S. Ghosh, A. Raychowdhury and S. Sen, ``X-DeepSCA: Cross-Device Deep Learning Side Channel Attack," 2019 56th ACM/IEEE Design Automation Conference (DAC), 2019, pp. 1-6, doi: 10.1145/3316781.3317934.
\bibitem{yt_fix}
``6 Ways to Fix Broken Wires - Beginner Through Pro," www.youtube.com, https://www.youtube.com/watch?v=xMqt\\5AoWPyI\&ab\_channel=LRN2DIY (accessed Aug. 23, 2025).
\bibitem{fix1}
D. Farquhar, ``How to fix a broken wire,'' The Silicon Underground, Dec. 04, 2019. https://dfarq.homeip.net/how-to-fix-a-broken-wire/ (accessed Aug. 23, 2025).
\bibitem{butt_con}
``How to Strip and Connect Wires With a Butt Connector,'' Instructables. https://www.instructables.com/How-to-strip-and-connect-wires-with-a-butt-connect/ (accessed Aug. 23, 2025).
\bibitem{ps2_1}
M. Vuagnoux and S. Pasini, ``Compromising electromagnetic emanations of wired and wireless keyboards,'' In USENIX security symposium, 2009, Vol. 8, pp. 1-16.
\bibitem{ps2_2}
L. Wang and B. Yu, ``Analysis and Measurement on the Electromagnetic Compromising Emanations of Computer Keyboards," 2011 Seventh International Conference on Computational Intelligence and Security, Sanya, China, 2011, pp. 640-643, doi: 10.1109/CIS.2011.146.
\bibitem{usb1_1}
A. Boitan, R. Bărtușică, S. Halunga, M. Popescu, \& I. Ionuță, ``Compromising electromagnetic emanations of wired USB keyboards,'' In Future Access Enablers for Ubiquitous and Intelligent Infrastructures: Third International Conference, FABULOUS 2017, Bucharest, Romania, October 12-14, 2017, Proceedings 3 (pp. 39-44). Springer International Publishing.
\bibitem{usb1_2}
D. Sim, H. S. Lee, J. Yook and K. Sim, ``Measurement and analysis of the compromising electromagnetic emanations from USB keyboard," 2016 Asia-Pacific International Symposium on Electromagnetic Compatibility (APEMC), Shenzhen, 2016, pp. 518-520, doi: 10.1109/APEMC.2016.7522785.
\bibitem{usb1_3}
H. -J. Choi, H. S. Lee, D. Sim, J. -G. Yook and K. Sim, ``Reconstruction of leaked signal from USB keyboards," 2016 URSI Asia-Pacific Radio Science Conference (URSI AP-RASC), Seoul, Korea (South), 2016, pp. 1281-1283, doi: 10.1109/URSIAP-RASC.2016.7601331.
\bibitem{nato_tempest}
I. Shopina, D. Khomiakov, N. Khrystynchenko, S. Zhukov, and D. Shpenov, ``CYBERSECURITY: LEGAL AND ORGANIZATIONAL SUPPORT IN LEADING COUNTRIES, NATO AND EU STANDARDS,'' Journal of Security \& Sustainability Issues, 9(3).
\bibitem{ce_wired_usb}
A. Boitan, R. Bărtușică, S. Halunga, M. Popescu and I. Ionuță, ``Compromising electromagnetic emanations of wired USB keyboards,'' In Future Access Enablers for Ubiquitous and Intelligent Infrastructures: Third International Conference, FABULOUS 2017, Bucharest, Romania, October 12-14, 2017, Proceedings 3 (pp. 39-44). Springer International Publishing.
\bibitem{van_eck}
W. Van Eck, ``Electromagnetic radiation from video display units: An eavesdropping risk?," Computers \& Security 4, no. 4 ,1985, 269-286.
\bibitem{wiki_temp}
``Tempest (codename)," Wikipedia. https://en.wikipedia.org/\\wiki/Tempest\_(codename). (accessed Aug. 23, 2025).
\bibitem{tech_rep_kuhn}
M. G. Kuhn, ``Compromising emanations: eavesdropping risks of computer displays," (Doctoral dissertation, University of Cambridge), 2002.
\bibitem{usb_stick}
B. Liu, Y. Xu, W. Huang and S. Guo, ``Detecting USB Storage Device Behaviors by Exploiting Electromagnetic Emanations," ICC 2022 - IEEE International Conference on Communications, 2022, pp. 4980-4985, doi: 10.1109/ICC45855.2022.9839155.
\bibitem{dnn_eman}
S. Liang, Z. Zhan, F. Yao, L. Cheng and Z. Zhang, ``Clairvoyance: Exploiting Far-field EM Emanations of GPU to ``See" Your DNN Models through Obstacles at a Distance," 2022 IEEE Security and Privacy Workshops (SPW), 2022, pp. 312-322, doi: 10.1109/SPW54247.2022.9833894.
\bibitem{smartphone}
B. B. Yilmaz, E. Mert Ugurlu, A. Zajić and M. Prvulovic, ``Cell-Phone Classification: A Convolutional Neural Network Approach Exploiting Electromagnetic Emanations," ICASSP 2020 - 2020 IEEE International Conference on Acoustics, Speech and Signal Processing (ICASSP), 2020, pp. 2862-2866, doi: 10.1109/ICASSP40776.2020.9054006.
\bibitem{bari_date23}
M. F. Bari, M. R. Chowdhury and S. Sen, ``Long Range Detection of Emanation from HDMI Cables Using CNN and Transfer Learning," 2023 Design, Automation \& Test in Europe Conference \& Exhibition (DATE), Antwerp, Belgium, 2023, pp. 1-6, doi: 10.23919/DATE56975.2023.10137263.
\bibitem{ims2022_eman} 
M. F. Bari, M. R. Chowdhury, B. Chatterjee and S. Sen, ``Detection of Rogue Devices using Unintended Near and Far-field Emanations with Spectral and Temporal Signatures," in IEEE/MTT-S International Microwave Symposium - IMS, 2022, pp. 591-594, doi: 10.1109/IMS37962.2022.9865347.
\bibitem{arduino_covert}
M. Hegarty, Y. E. Sagduyu, T. Erpek and Y. Shi, ``Deep Learning for Spectrum Awareness and Covert Communications via Unintended RF Emanations." In Proceedings of the 2022 ACM Workshop on Wireless Security and Machine Learning, pp. 27-32. 2022.
\bibitem{gsmem}
M. Guri, A. Kachlon, O. Hasson, G. Kedma, Y. Mirsky, \& Y. Elovici, ``GSMem: Data exfiltration from Air-Gapped computers over GSM frequencies," In 24th USENIX Security Symposium (USENIX Security 15) (pp. 849-864).
\bibitem{airfi}
M. Guri, ``AIR-FI: Leaking Data From Air-Gapped Computers Using Wi-Fi Frequencies," in IEEE Transactions on Dependable and Secure Computing, vol. 20, no. 3, pp. 2547-2564, 1 May-June 2023, doi: 10.1109/TDSC.2022.3186627.
\bibitem{bitjabber}
Z. Zhan, Z. Zhang and X. Koutsoukos, ``BitJabber: The World's Fastest Electromagnetic Covert Channel," 2020 IEEE International Symposium on Hardware Oriented Security and Trust (HOST), San Jose, CA, USA, 2020, pp. 35-45, doi: 10.1109/HOST45689.2020.9300268.
\bibitem{noisehopper}
M. F. Bari and S. Sen, ``NoiseHopper: Emission Hopping Air-Gap Covert Side Channel with Lower Probability of Detection," 2024 IEEE International Symposium on Hardware Oriented Security and Trust (HOST), Tysons Corner, VA, USA, 2024, pp. 21-32, doi: 10.1109/HOST55342.2024.10545402.
\bibitem{touch_access}
Y. Liu, Z. Xu, Z. Qin, L. Ou and W. Jin, ``TouchAccess: Unlock IoT Devices on Touching by Leveraging Human-Induced EM Emanations," in IEEE Transactions on Mobile Computing, doi: 10.1109/TMC.2024.3414992.
\bibitem{bari_dirac}
M. F. Bari, B. Chatterjee and S. Sen, ``DIRAC: Dynamic-IRregulAr Clustering Algorithm with Incremental Learning for RF-Based Trust Augmentation in IoT Device Authentication," 2021 IEEE International Symposium on Circuits and Systems (ISCAS), Daegu, Korea, 2021, pp. 1-5, doi: 10.1109/ISCAS51556.2021.9401403.
\bibitem{bari_rfpuf_ims}
M. F. Bari, B. Chatterjee, K. Sivanesan, L. L. Yang and S. Sen, ``High Accuracy RF-PUF for EM Security through Physical Feature Assistance using Public Wi-Fi Dataset," 2021 IEEE MTT-S International Microwave Symposium (IMS), Atlanta, GA, USA, 2021, pp. 108-111, doi: 10.1109/IMS19712.2021.9574917.
\bibitem{bari_rfpuf_fe}
M. F. Bari, P. Agrawal, B. Chatterjee, and S. Sen, ``Statistical Analysis Based Feature Selection Enhanced RF-PUF With $>$99.8\% Accuracy on Unmodified Commodity Transmitters for IoT Physical Security," in Frontiers in Electronics, vol. 3, Apr. 2022, doi: https://doi.org/10.3389/felec.2022.856284.
\bibitem{eman_finger}
J. Feng et al., ``Fingerprinting IoT Devices Using Latent Physical Side-Channels,'' Proceedings of the ACM on interactive, mobile, wearable and ubiquitous technologies, vol. 7, no. 2, pp. 1–26, Jun. 2023, doi: https://doi.org/10.1145/3596247.
\bibitem{eman_forensic}
A. P. Sayakkara and N. -A. Le-Khac, ``Electromagnetic Side-Channel Analysis for IoT Forensics: Challenges, Framework, and Datasets," in IEEE Access, vol. 9, pp. 113585-113598, 2021, doi: 10.1109/ACCESS.2021.3104525.
\bibitem{eman_mb}
E. J. Jorgensen, F. T. Werner, M. Prvulovic, and A. Zajić, ``Deep Learning Classification of Motherboard Components by Leveraging EM Side-Channel Signals,'' Journal of Hardware and Systems Security, vol. 5, no. 2, pp. 114–126, Jun. 2021, doi: https://doi.org/10.1007/s41635-021-00116-2.
\bibitem{bari_iotj}
M. F. Bari, M. R. Chowdhury and S. Sen, ``A Computational Harmonic Detection Algorithm to Detect Data Leakage through EM Emanation," in IEEE Internet of Things Journal, doi: 10.1109/JIOT.2025.3578511.
\bibitem{akb1}
T. Zhu, Q. Ma, S. Zhang and Y. Liu, ``Context-free Attacks Using Keyboard Acoustic Emanations,'' In Proceedings of the 2014 ACM SIGSAC Conference on Computer and Communications Security (CCS '14). Association for Computing Machinery, New York, NY, USA, 453–464. https://doi.org/10.1145/2660267.2660296
\bibitem{akb2}
L. Zhuang, F. Zhou and J. D. Tygar, ``Keyboard acoustic emanations revisited,'' ACM Trans. Inf. Syst. Secur. 13, 1, Article 3 (October 2009), 26 pages. https://doi.org/10.1145/1609956.1609959
\bibitem{akb3}
D. Slater, S. Novotney, J. Moore, S. Morgan and S. Tenaglia, ``Robust keystroke transcription from the acoustic side-channel,'' In Proceedings of the 35th Annual Computer Security Applications Conference (ACSAC '19). Association for Computing Machinery, New York, NY, USA, 776–787. https://doi.org/10.1145/3359789.3359816
\bibitem{akb4}
J. Liu, Y. Wang, G. Kar, Y. Chen, J. Yang and M. Gruteser, ``Snooping Keystrokes with mm-level Audio Ranging on a Single Phone,'' In Proceedings of the 21st Annual International Conference on Mobile Computing and Networking (MobiCom '15). Association for Computing Machinery, New York, NY, USA, 142–154. https://doi.org/10.1145/2789168.2790122
\bibitem{akb5}
T. Halevi and N. Saxena, ``A closer look at keyboard acoustic emanations: random passwords, typing styles and decoding techniques,'' In Proceedings of the 7th ACM Symposium on Information, Computer and Communications Security (ASIACCS '12). Association for Computing Machinery, New York, NY, USA, 89–90. https://doi.org/10.1145/2414456.2414509
\bibitem{thermal_kb}
T. Kaczmarek, E. Ozturk and G. Tsudik, ``Thermanator: Thermal Residue-Based Post Factum Attacks on Keyboard Data Entry,'' In Proceedings of the 2019 ACM Asia Conference on Computer and Communications Security (Asia CCS '19). Association for Computing Machinery, New York, NY, USA, 586–593. https://doi.org/10.1145/3321705.3329846
\bibitem{kb_motion}
L. Cai and H. Chen, ``On the practicality of motion based keystroke inference attack,'' In International Conference on Trust and Trustworthy Computing (pp. 273-290). Berlin, Heidelberg: Springer Berlin Heidelberg.
\bibitem{kb_wifi}
K. Ali, A. X. Liu, W. Wang and M. Shahzad, ``Keystroke Recognition Using WiFi Signals,'' In Proceedings of the 21st Annual International Conference on Mobile Computing and Networking (MobiCom '15). Association for Computing Machinery, New York, NY, USA, 90–102. https://doi.org/10.1145/2789168.2790109
\bibitem{periscope}
W. Jin, S. Murali, H. Zhu and M. Li, ``Periscope: A Keystroke Inference Attack Using Human Coupled Electromagnetic Emanations,'' In Proceedings of the 2021 ACM SIGSAC Conference on Computer and Communications Security (CCS '21). Association for Computing Machinery, New York, NY, USA, 700–714. https://doi.org/10.1145/3460120.3484549
\bibitem{th_kb_1}
R. I. Sokolov, R. R. Abdullin and D. A. Dolmatov, ``Development of synchronization system for signal reception and recovery from USB-keyboard compromising emanations," 2016 2nd International Conference on Industrial Engineering, Applications and Manufacturing (ICIEAM), Chelyabinsk, Russia, 2016, pp. 1-4, doi: 10.1109/ICIEAM.2016.7911553.
\bibitem{th_kb_2}
R. I. Sokolov and R. R. Abdullin, ``Determined factor parameter analysis for system of information recovery from USB-keyboard compromising emanations," 2017 International Applied Computational Electromagnetics Society Symposium - Italy (ACES), Firenze, Italy, 2017, pp. 1-2, doi: 10.23919/ROPACES.2017.7916332.
\bibitem{usb2protocol}
``Universal Serial Bus Specification Compaq Hewlett-Packard Intel Lucent Microsoft NEC Philips Revision 2.0,'' 2000. Available: https://eater.net/downloads/usb\_20.pdf. (accessed June 19, 2025).


\end{thebibliography}



\end{document}